\documentclass{article}%
\usepackage{amsmath}
\usepackage{amsfonts}
\usepackage{amssymb}
\usepackage{graphicx}%
\providecommand{\U}[1]{\protect\rule{.1in}{.1in}}
\begin{document}

\author{Giuseppe Castagnoli
\and Pieve Ligure, Italy, giuseppe.castagnoli@gmail.com}
\title{An explanation of the quantum speed up}
\maketitle

\begin{abstract}
In former work, we showed that a quantum algorithm requires the number of
operations (oracle's queries) of a classical algorithm that knows in advance
50\% of the information that specifies the solution of the problem. We gave a
preliminary theoretical justification of this "50\% rule" and checked that the
rule holds for a variety of quantum algorithms. Now, we make explicit the
information about the solution available to the algorithm throughout the
computation. The final projection on the solution becomes acquisition of the
knowledge of the solution on the part of the algorithm. Backdating to before
running the algorithm a time-symmetric part of this projection, feeds back to
the input of the computation 50\% of the information acquired by reading the solution.

\end{abstract}

\section{Introduction}

We provide an explanation of the quantum speed up -- the fact that quantum
problem solving requires fewer operations than its classical counterpart.

Let us consider, for example, the quantum data base search problem. We resort
to a visualization. With data base size $N=4$, we have a chest of 4 drawers,
numbered 00, 01, 10, and 11, a ball, and two players. The first player, the
oracle, hides the ball in drawer number $\mathbf{k}\equiv\mathbf{~}k_{0}%
,k_{1}$ and gives to the second player, Alice -- the algorithm that solves the
problem -- the chest of drawers. This is represented by a black box that,
given in input a drawer number $\mathbf{x}\equiv x_{0},x_{1}$, computes the
Kronecker function $f_{\mathbf{k}}\left(  \mathbf{x}\right)  =\delta\left(
\mathbf{k},\mathbf{x}\right)  $ (which is 1 if $\mathbf{k}=\mathbf{x}$, 0
otherwise). Alice should find the number of the drawer with the ball, and this
is done by computing $\delta\left(  \mathbf{k},\mathbf{x}\right)  $ for
different values of $\mathbf{x}$ -- by opening different drawers (we also say:
by \textit{querying the oracle} with different values of $\mathbf{x}$). A
classical algorithm requires 2.25 computations of $\delta\left(
\mathbf{k},\mathbf{x}\right)  $\ on average, 3 computations if one wants to be
a priori certain of finding the solution. The quantum algorithm yields the
solution with certainty with just one computation.

The reason for the quantum speed up is not well understood. For example,
recently Gross et al. $\left[  1\right]  $ asserted that the exact reason for
it has never been pinpointed.

Let $\mathcal{I}$\ be the information acquired by reading the solution
produced by the quantum algorithm. The explanation we gave in Ref. $\left[
2\right]  $ and $\left[  3\right]  $ is that a quantum algorithm requires
fewer oracle's queries\ because it knows in advance 50\% of $\mathcal{I}$. The
computation performed by the quantum algorithm is a superposition of all the
computation histories of a classical algorithm that knows in advance 50\% of
$\mathcal{I}$. Each history corresponds to a possible way of taking 50\% of
$\mathcal{I}$ and a possible result of computing the missing information. We
called this the "50\% rule".

A key step was extending the representation of the quantum algorithm to the
random generation of a value of $\mathbf{k}$. Consequently, the end of the
unitary part of the algorithm is a state of maximal entanglement between
oracle's choice $\mathbf{k}$\ and solution of the problem. Measuring the
content of the computer registers produces both a value of $\mathbf{k}$\ at
random and the corresponding solution -- namely the value of $\mathbf{x}%
$\ such that $\delta\left(  \mathbf{k},\mathbf{x}\right)  =1$. The 50\% rule
is derived from the assumption that causality between these two correlated
measurement outcomes is \textquotedblleft mutual and
symmetrical\textquotedblright\ -- see also Ref. $\left[  4\right]  $.

In this paper, we represent the mutual determination between oracle's choice
and solution in a different way. The quantum algorithm is seen under two
time-symmetric perspectives. In \textit{Alice's perspective},\ given the value
of $\mathbf{k}$, the algorithm finds the value of $\mathbf{x}$\ such that
$\delta\left(  \mathbf{k},\mathbf{x}\right)  =1$. In the time-symmetric
\textit{oracle's perspective}, the input and the output of the computation are
exchanged: given the value of $\mathbf{x}$, the algorithm finds the value of
$\mathbf{k}$ such that $\delta\left(  \mathbf{k},\mathbf{x}\right)
=1$\textit{.}

Furthermore, these two perspectives are represented in a relational way. In
Alice's perspective the observer is Alice, in the oracle's perspective the
observer is the oracle. In the relational representation in Alice's (the
oracle's)\ perspective, we make explicit the information about the solution
available to Alice (the oracle) throughout the computation. The final
projection on the solution becomes acquisition of the knowledge of the
solution on the part of Alice (the oracle).

In either perspective, backdating to before running the algorithm a percentage
of the final projection on the solution, makes available at the input of the
computation the same percentage of $\mathcal{I}$. Since the two perspectives
must yield the same speed up, the projection on the solution must share out
evenly between the two perspectives. This means that, in either perspective,
the quantum algorithm knows in advance 50\% of $\mathcal{I}$.

The potential applications of the 50\% rule look interesting. According to the
rule, the quantum speed up in terms of number of oracle's queries comes from
comparing two classical algorithms, with and without advanced information.
Therefore the rule could be used for a characterization of the problems
solvable with a quantum speed up in an entirely computer science framework,
with no physics involved.

Section 2 provides the new derivation of the 50\% rule in the case of Grover's
algorithm. In section 3, we check that the rule holds for a class of quantum
algorithms that yield an exponential speed up. In section 4, by way of
exemplification, we develop a new quantum algorithm out of the 50\% rule. In
section 5 we draw the conclusions.

\section{Explaining the quantum speed up}

We explain the speed up on Grover's $\left[  5\right]  $ quantum data base
search algorithm, first for data base size $N=4$, then for $N>4$.

\subsection{Reviewing the extended representation of Grover's algorithm}

In Ref. $\left[  2\right]  $, we extended the representation of Grover's
algorithm to the random generation of a value of $\mathbf{k}$. We review this
representation, which is the kernel of the relational representations of the
quantum algorithm developed in the next section.

We have three computer registers: (i) a two-qubit register $X$ contains the
argument $\mathbf{x}$ to query the oracle with (namely the input of the
computation of $\delta\left(  \mathbf{k},\mathbf{x}\right)  $), (ii) a
one-qubit register $V$ is meant to contain the result of the computation,
modulo 2 added to its initial content for logical reversibility, and (iii) a
two-qubit register $K$ (just a conceptual reference)\ contains the oracle's
choice of a value of $\mathbf{k}$.

The initial state of the three registers is:%

\begin{align}
&  \frac{1}{4\sqrt{2}}\left(  \left\vert 00\right\rangle _{K}+\left\vert
01\right\rangle _{K}+\left\vert 10\right\rangle _{K}+\left\vert
11\right\rangle _{K}\right)  \left(  \left\vert 00\right\rangle _{X}%
+\left\vert 01\right\rangle _{X}+\left\vert 10\right\rangle _{X}+\left\vert
11\right\rangle _{X}\right) \label{input}\\
&  \left(  \left\vert 0\right\rangle _{V}-\left\vert 1\right\rangle
_{V}\right)  .\nonumber
\end{align}

The computation of $\delta\left(  \mathbf{k},\mathbf{x}\right)  $ is performed
in quantum parallelism on each term of the superposition. For example, the
input term $-\left\vert 01\right\rangle _{K}\left\vert 01\right\rangle
_{X}\left\vert 1\right\rangle _{V}$ means that the input of the black box is
$\mathbf{k}=01,$ $\mathbf{x}=01$ and that the initial content of register
$V$\ is 1. The computation yields $\delta\left(  01,01\right)  =1$, which
modulo 2 added to the initial content of $V$\ yields the output term
$-\left\vert 01\right\rangle _{K}\left\vert 01\right\rangle _{X}\left\vert
0\right\rangle _{V}$ ($K$\ and $X$ keep the memory of the input for logical
reversibility). In the overall, the computation of $\delta\left(
\mathbf{k},\mathbf{x}\right)  $ sends state (\ref{input}) into:%

\begin{equation}
\frac{1}{4\sqrt{2}}\left[
\begin{array}
[c]{c}%
\left\vert 00\right\rangle _{K}\left(  -\left\vert 00\right\rangle
_{X}+\left\vert 01\right\rangle _{X}+\left\vert 10\right\rangle _{X}%
+\left\vert 11\right\rangle _{X}\right)  +\\
\left\vert 01\right\rangle _{K}\left(  \left\vert 00\right\rangle
_{X}-\left\vert 01\right\rangle _{X}+\left\vert 10\right\rangle _{X}%
+\left\vert 11\right\rangle _{X}\right)  +\\
\left\vert 10\right\rangle _{K}\left(  \left\vert 00\right\rangle
_{X}+\left\vert 01\right\rangle _{X}-\left\vert 10\right\rangle _{X}%
+\left\vert 11\right\rangle _{X}\right)  +\\
\left\vert 11\right\rangle _{K}\left(  \left\vert 00\right\rangle
_{X}+\left\vert 01\right\rangle _{X}+\left\vert 10\right\rangle _{X}%
-\left\vert 11\right\rangle _{X}\right)
\end{array}
\right]  \left(  \left\vert 0\right\rangle _{V}-\left\vert 1\right\rangle
_{V}\right)  , \label{secondstage}%
\end{equation}
a maximally entangled state where four orthogonal states of $K$, each
corresponding to a single value of $\mathbf{k}$, are correlated with four
orthogonal states of $X$. This means that the information about the value of
$\mathbf{k}$\ has propagated to $X$.

A suitable rotation of the\ measurement basis of $X$ (or $K$ -- see further
below), transforms entanglement between $K$\ and $X$\ into correlation between
the outcomes of measuring their contents, sending state (\ref{secondstage}) into:%

\begin{align}
&  \frac{1}{\sqrt{2}}\left(  \left\vert 00\right\rangle _{K}\left\vert
00\right\rangle _{X}+\left\vert 01\right\rangle _{K}\left\vert 01\right\rangle
_{X}+\left\vert 10\right\rangle _{K}\left\vert 10\right\rangle _{X}+\left\vert
11\right\rangle _{K}\left\vert 11\right\rangle _{X}\right) \label{output}\\
&  \left(  \left\vert 0\right\rangle _{V}-\left\vert 1\right\rangle
_{V}\right)  .\nonumber
\end{align}

Let us call $\left[  X\right]  $ the content of register $X$\ and $\left[
K\right]  $ the content of register $K$. Measuring $\left[  X\right]  $ and/or
$\left[  K\right]  $\ in the output state (\ref{output}) yields the solution
$\mathbf{x}=\mathbf{k}$, also generating the oracle's choice at random. In
fact, since the algorithm is the identity in the Hilbert space of register $K$
(having rotated the basis of register $X$), nothing changes if we measure
$\left[  K\right]  $ in the initial state (\ref{input}) -- which generates the
oracle's choice at random -- and $\left[  X\right]  $ in the output state
(\ref{output}).

Until now we have seen Grover's algorithm in "Alice's perspective": the oracle
chooses a value of $\mathbf{k}$ at random and Alice finds it, first by
computing $\delta\left(  \mathbf{k},\mathbf{x}\right)  $\ for all the values
of $\mathbf{x}$\ in quantum superposition, then by rotating the basis of
register $X$, and finally by measuring $\left[  X\right]  $. In the
time-symmetric oracle's perspective, Alice chooses a value of $\mathbf{x}$ at
random and the oracle finds it, first by computing $\delta\left(
\mathbf{k},\mathbf{x}\right)  $\ for all the values of $\mathbf{k}$\ in
quantum superposition, then by rotating the basis of register $K$, and finally
by measuring $\left[  K\right]  $. We note that states (\ref{input}) through
(\ref{output}) are common to the two perspectives.

\subsection{Relational representation of the quantum algorithm}

We develop two relational $\left[  6\right]  $ representations of the quantum
algorithm. In one the observer is Alice, who is in control of register $X$
(from now on, by \textit{Alice's perspective} we mean the relational
representation of the quantum algorithm with respect to Alice). In the other
the observer is the oracle, who is in control of register $K$ (this is
\textit{the oracle's perspective}).

By definition, initially Alice does not know the oracle's choice -- to fix
ideas, say that this choice (which does not need to be random) is
$\mathbf{k}=00$. This does not affect the initial state of the algorithm in
Alice's perspective, which is anyhow:%

\begin{align}
&  \frac{1}{4\sqrt{2}}\left(  \left\vert 00\right\rangle _{K}+\operatorname{e}%
^{i\varphi_{01}}\left\vert 01\right\rangle _{K}+\operatorname{e}%
^{i\varphi_{10}}\left\vert 10\right\rangle _{K}+\operatorname{e}%
^{i\varphi_{11}}\left\vert 11\right\rangle _{K}\right) \label{inputa}\\
&  \left(  \left\vert 00\right\rangle _{X}+\left\vert 01\right\rangle
_{X}+\left\vert 10\right\rangle _{X}+\left\vert 11\right\rangle _{X}\right)
\left(  \left\vert 0\right\rangle _{V}-\left\vert 1\right\rangle _{V}\right)
,\nonumber
\end{align}
where the $\varphi_{ij}$ are random phases with uniform distribution in
$\left[  0,2\pi\right]  $. We use the random phase representation of density
operators to keep the state vector representation of the quantum algorithm
(the density operator is the average over $\varphi$ of the product of the ket
by the bra). Register $K$\ is in a maximally mixed state. The two bits von
Neumann entropy of the state of register $K$\ represents Alice's initial
ignorance of the oracle's choice.

In Alice's perspective, the algorithm is the identity in the Hilbert space of
register $K$, thus state (\ref{inputa}) develops like state (\ref{input}).
Computing \ $\delta$\ and rotating the basis of register $X$ sends
(\ref{inputa}) into the \textit{output state}:%

\begin{align}
&  \frac{1}{2\sqrt{2}}\left(  \left\vert 00\right\rangle _{K}\left\vert
00\right\rangle _{X}+\operatorname{e}^{i\varphi_{01}}\left\vert
01\right\rangle _{K}\left\vert 01\right\rangle _{X}+\operatorname{e}%
^{i\varphi_{10}}\left\vert 10\right\rangle _{K}\left\vert 10\right\rangle
_{X}+\operatorname{e}^{i\varphi_{11}}\left\vert 11\right\rangle _{K}\left\vert
11\right\rangle _{X}\right) \label{outputa}\\
&  \left(  \left\vert 0\right\rangle _{V}-\left\vert 1\right\rangle
_{V}\right)  .\nonumber
\end{align}

The measurement of $\left[  X\right]  $ (or, indifferently, $\left[  K\right]
$) projects the output state on%

\begin{equation}
\frac{1}{\sqrt{2}}\left\vert 00\right\rangle _{K}\left\vert 00\right\rangle
_{X}\left(  \left\vert 0\right\rangle _{V}-\left\vert 1\right\rangle
_{V}\right)  \label{afme}%
\end{equation}
and yields the solution $\mathbf{x}=\mathbf{k}=00$. This projection is random
to Alice but, seen from outside, it occurs on the eigenstate of the eigenvalue
of $\mathbf{k}$ chosen by the oracle ($\mathbf{k}=$ $00$). It is reduction of
ignorance of the solution on the part of Alice -- her acquisition of the
knowledge of the oracle's choice. Correspondingly, the entropy decreases from
the two bits of states (\ref{inputa}) and (\ref{outputa}) to the value zero of
state (\ref{afme}).

In the oracle's perspective, the initial state of the algorithm is:%

\begin{align}
&  \frac{1}{4\sqrt{2}}\left(  \left\vert 00\right\rangle _{K}+\left\vert
01\right\rangle _{K}+\left\vert 10\right\rangle _{K}+\left\vert
11\right\rangle _{K}\right) \label{oracle}\\
&  \left(  \left\vert 00\right\rangle _{X}+\operatorname{e}^{i\varphi_{01}%
}\left\vert 01\right\rangle _{X}+\operatorname{e}^{i\varphi_{10}}\left\vert
10\right\rangle _{X}+\operatorname{e}^{i\varphi_{11}}\left\vert
11\right\rangle _{X}\right)  \left(  \left\vert 0\right\rangle _{V}-\left\vert
1\right\rangle _{V}\right)  .\nonumber
\end{align}
\ Note that the two perspectives transforms into one another by swapping the
labels $K$ and $X$. Computing \ $\delta$\ and rotating the basis of register
$K$ sends state (\ref{oracle}) into the output state (\ref{outputa}),
invariant under the swapping of $K$\ and $X$ and common to both perspectives.

\subsection{Taking advantage of backdated projection on the solution}

Let us break down $\left[  X\right]  $, the content of register $X$, into
content of first qubit $\left[  X_{0}\right]  $ and content of second qubit
$\left[  X_{1}\right]  $ (other ways of halving $\left[  X\right]  $ will be
considered in the next section). Let $x_{0}$ ($x_{1}$)\ be the eigenvalue
obtained by measuring $\left[  X_{0}\right]  $ ($\left[  X_{1}\right]  $) in
the output state (\ref{outputa}). We define in a similar way $\left[
K_{0}\right]  $, $\left[  K_{1}\right]  $, $k_{0}$, and $k_{1}$.\ In the
assumption that $\mathbf{x}=\mathbf{k}=00$, measuring, say, $\left[
X_{0}\right]  $ in (\ref{outputa})\ yields $x_{0}=k_{0}=0$, measuring $\left[
K_{1}\right]  $ yields $x_{1}=k_{1}=0$. Together, the two measurements project
the output state (\ref{outputa}) on the eigenstate corresponding to the solution.

The measurement of $\left[  X_{0}\right]  $ (or, indifferently, $\left[
K_{0}\right]  $) projects the output state on:%

\begin{equation}
\frac{1}{2}\left(  \left\vert 00\right\rangle _{K}\left\vert 00\right\rangle
_{X}+\operatorname{e}^{i\varphi_{01}}\left\vert 01\right\rangle _{K}\left\vert
01\right\rangle _{X}\right)  \left(  \left\vert 0\right\rangle _{V}-\left\vert
1\right\rangle _{V}\right)  . \label{poutputa}%
\end{equation}
We backdate the related projection along Alice's perspective. This projects
the initial state of Alice's perspective, (\ref{inputa}), on:%
\begin{equation}
\frac{1}{4}\left(  \left\vert 00\right\rangle _{K}+\operatorname{e}%
^{i\varphi_{01}}\left\vert 01\right\rangle _{K}\right)  \left(  \left\vert
00\right\rangle _{X}+\left\vert 01\right\rangle _{X}+\left\vert
10\right\rangle _{X}+\left\vert 11\right\rangle _{X}\right)  \left(
\left\vert 0\right\rangle _{V}-\left\vert 1\right\rangle _{V}\right)  .
\label{ai}%
\end{equation}
We have applied to state (\ref{poutputa}) the inverse of the time forward
unitary transformation (the result is evident keeping in mind that, in Alice's
perspective, the algorithm is the identity in the Hilbert space of register
$K$). State (\ref{ai}) says that, "after" the backdated projection, Alice
knows before running the algorithm that the oracle's choice is either
$\mathbf{k}=00$ or $\mathbf{k}=01$, namely that $k_{0}=0$. Correspondingly,
the entropy representing Alice's ignorance of the solution has decreased from
two to one bit.

The measurement of $\left[  K_{1}\right]  $ (or, indifferently, $\left[
X_{1}\right]  $)\ projects the output state on:%

\begin{equation}
\frac{1}{2}\left(  \left\vert 00\right\rangle _{K}\left\vert 00\right\rangle
_{X}+\operatorname{e}^{i\varphi_{10}}\left\vert 10\right\rangle _{K}\left\vert
10\right\rangle _{X}\right)  \left(  \left\vert 0\right\rangle _{V}-\left\vert
1\right\rangle _{V}\right)  . \label{poutputo}%
\end{equation}
We backdate this projection along the oracle's perspective. This projects the
initial state of the oracle's perspective, (\ref{oracle}), on:%
\begin{equation}
\frac{1}{4}\left(  \left\vert 00\right\rangle _{K}+\left\vert 01\right\rangle
_{K}+\left\vert 10\right\rangle _{K}+\left\vert 11\right\rangle _{K}\right)
\left(  \left\vert 00\right\rangle _{X}+\operatorname{e}^{i\varphi_{10}%
}\left\vert 10\right\rangle _{X}\right)  \left(  \left\vert 0\right\rangle
_{V}-\left\vert 1\right\rangle _{V}\right)  \label{oi}%
\end{equation}
(now the algorithm is the identity in the Hilbert space of $X$). This means
that the oracle knows in advance that $x_{1}=k_{1}=0$, in fact the other bit
of information about the solution.

This is what is needed, because the amount of advanced knowledge of the
solution should be the same in either perspective (the two perspectives must
yield the same speed up).

We discuss the other ways of backdating the projection on the solution.
Backdating along Alice's perspective the projection due to measuring $\left[
X_{1}\right]  $\ and along the oracle's perspective the projection due to
measuring $\left[  K_{0}\right]  $, just exchanges the two bits known in
advance by respectively Alice and the oracle. Backdating along Alice's (the
oracle's) perspective the entire projection on the solution, would leave
nothing to backdate along the oracle's (Alice's) perspective, in conflict with
the requirement that the projection on the solution shares out evenly between
the two perspectives.

The above shows that: (i) the advanced knowledge of 50\% of $\mathcal{I}$,
highlighted in Ref. $\left[  2\right]  $, is backdated projection on the
solution and (ii) the algorithm cannot exploit in advance more than 50\% of
the projection on the solution.

In view of what will follow, we highlight another way of seeing Alice's
advanced knowledge. In Alice's perspective, the reduced density operator of
register $K$ is $\frac{1}{2}\left(  \left\vert 00\right\rangle _{K}%
+\operatorname{e}^{i\varphi_{01}}\left\vert 01\right\rangle _{K}%
+\operatorname{e}^{i\varphi_{10}}\left\vert 10\right\rangle _{K}%
+\operatorname{e}^{i\varphi_{11}}\left\vert 11\right\rangle _{K}\right)  $
throughout the unitary part of the quantum algorithm (which is the identity on
this operator). The measurement of $\left[  X_{0}\right]  $ (or $\left[
K_{0}\right]  $)\ in the output state (\ref{outputa}) projects it on $\frac
{1}{\sqrt{2}}\left(  \left\vert 00\right\rangle _{K}+\operatorname{e}%
^{i\varphi_{01}}\left\vert 01\right\rangle _{K}\right)  $. This projection
goes back unaltered to before running the algorithm, yielding Alice's advanced knowledge.

\subsection{Superposition of the histories}

We show how the quantum algorithm exploits the advanced knowledge of the
solution. We put ourselves in Alice's perspective. Until now we have
considered two ways of halving the projection on the solution: measuring the
binary observable $\left[  X_{0}\right]  $, which tells whether the oracle's
choice $\mathbf{k}$ belongs to $\left\{  00,01\right\}  $ or $\left\{
10,11\right\}  $, or measuring $\left[  X_{1}\right]  $, which tells whether
$\mathbf{k}$ belongs to $\left\{  00,10\right\}  $ or $\left\{  01,11\right\}
$. There is a third binary observable, say $\left[  X_{+}\right]  $, whose
measurement tells whether $\mathbf{k}$ belongs to $\left\{  00,11\right\}  $
or $\left\{  01,10\right\}  $. Measuring any pair of these observables
projects the output state (\ref{outputa})\ on the solution. We note that the
results of the former section remain unaltered if we replace either $\left[
X_{0}\right]  $\ or $\left[  X_{1}\right]  $\ by $\left[  X_{+}\right]  $ and,
correspondingly, either $\left[  K_{0}\right]  $\ or $\left[  K_{1}\right]
$\ by $\left[  K_{+}\right]  $.

In the overall, there are $6$ halved projections, or ways of knowing in
advance one bit of $\mathcal{I}$: knowing that $\mathbf{k}$ belongs to
$\left\{  00,01\right\}  $, or $\left\{  10,11\right\}  $, ..., or $\left\{
01,10\right\}  $. Each halved projection originates 8 classical computation
histories, as follows.

We assume that Alice knows in advance that $\mathbf{k}$ belongs to $\left\{
00,01\right\}  $. To compute the missing bit, she should query the oracle with
either $\mathbf{x}=00$ or $\mathbf{x}=01$. We assume that she queries with
$\mathbf{x}=00$. If the outcome of the computation is $\delta=1$, this means
that $\mathbf{k}=00$. This originates two histories, depending on the initial
state of register $V$. History \# 1: initial state $\left\vert 00\right\rangle
_{K}\left\vert 00\right\rangle _{X}\left\vert 0\right\rangle _{V}$, state
after the computation $\left\vert 00\right\rangle _{K}\left\vert
00\right\rangle _{X}\left\vert 1\right\rangle _{V}$. History \# 2: initial
state $\left\vert 00\right\rangle _{K}\left\vert 00\right\rangle
_{X}\left\vert 1\right\rangle _{V}$, state after the computation $\left\vert
00\right\rangle _{K}\left\vert 00\right\rangle _{X}\left\vert 0\right\rangle
_{V}$.\ If the outcome of the computation is $\delta=0$, this means that
$\mathbf{k}=01$. This originates two histories, depending on the initial state
of register $V$. History \# 3: initial state $\left\vert 01\right\rangle
_{K}\left\vert 00\right\rangle _{X}\left\vert 0\right\rangle _{V}$, state
after the computation $\left\vert 01\right\rangle _{K}\left\vert
00\right\rangle _{X}\left\vert 0\right\rangle _{V}$. History \# 4: initial
state $\left\vert 01\right\rangle _{K}\left\vert 00\right\rangle
_{X}\left\vert 1\right\rangle _{V}$, state after the computation $\left\vert
01\right\rangle _{K}\left\vert 00\right\rangle _{X}\left\vert 1\right\rangle
_{V}$.\ If she queries the oracle with $\mathbf{x}=01$ instead, this
originates other 4 histories. Etc.

If we sum together all the different histories (some histories are originated
more than once), each with a suitable phase, and normalize, we obtain the
transformation of state (\ref{input})\ into (\ref{secondstage}), namely the
oracle's query stage of Grover's algorithm. For simplicity, we consider the
kernel of the quantum algorithm.

Rotating the basis of register $X$ transforms state (\ref{secondstage}) into
state (\ref{output}). Correspondingly, each classical history in quantum
notation branches into four histories, the branches of different histories
interfere with one another to give state (\ref{output}).

The 50\% rule only establishes that the quantum algorithm can be broken down
into a superposition of such histories, the history phases and the final
rotation of the basis of register $X$ are what is needed for the breaking down.

\subsection{Synthesizing the quantum algorithm out of the advanced information
classical algorithm}

As from Ref. $\left[  2\right]  $, the history phases that reconstruct the
quantum algorithm also maximize the entanglement between $K$ and $X$ after the
computation of $\delta$ -- see state (\ref{secondstage}). Then the rotation of
the basis of $X$ transforms this entanglement into correlation between the
outcomes of measuring $\left[  K\right]  $ and $\left[  X\right]  $ -- see
state (\ref{output}).

This, in principle, allows to synthesize the quantum algorithm out of the
advanced information classical algorithm. We should choose history phases and
rotation of the basis of $X$ in such a way that they maximize: (i) correlation
between the outcomes of measuring $\left[  K\right]  $ and $\left[  X\right]
$, or (ii) interference between histories, or (iii) the information about the
solution readable in $X$\ at the end of the algorithm.

\subsection{Generalizing to $N>4$}

We check that the explanation of the quantum speed up holds also for
$N=2^{n}>4$. Registers $K$\ and $X$\ are $n$-qubit each. Register $V$\ is
one-qubit. Given the advanced knowledge of $n/2$ bits, in order to compute the
missing $n/2$ bits we should compute $\delta\left(  \mathbf{k},\mathbf{x}%
\right)  $ for all the values of $\mathbf{x}$\ in quantum superposition\ and
rotate the basis of $X$ (in Alice's perspective) an\ $\operatorname{O}\left(
2^{n/2}\right)  $ times. The output state (\ref{output}) becomes:%

\begin{equation}
\frac{1}{2^{\left(  n+1\right)  /2}}\left(  \sum_{\mathbf{k}=0}^{2^{n}%
-1}\left\vert \mathbf{k}\right\rangle _{K}\left\vert \mathbf{k}\right\rangle
_{X}\right)  \left(  \left\vert 0\right\rangle _{V}-\left\vert 1\right\rangle
_{V}\right)  , \label{outgen}%
\end{equation}
we have considered for simplicity only the kernel of the quantum algorithm.
Measuring $\left[  X\right]  $ (or $\left[  K\right]  $) projects
(\ref{outgen}) on the solution. According to the rationale of section 2.3, we
should halve the final projection on the solution in all possible ways; for
example, by measuring $\left[  X_{0}\right]  ,~...,~\left[  X_{\frac{n}{2}%
-1}\right]  $ (or,\ indifferently, $\left[  K_{0}\right]  ,~...,~\left[
K_{\frac{n}{2}-1}\right]  $). Evidently, the considerations of section 2.3
apply also here: backdating a half projection in one perspective, makes
available at the input of the algorithm the corresponding 50\% of
$\mathcal{I}$. The other half projection (in the example, that due to
measuring $\left[  X_{\frac{n}{2}}\right]  ,~...,~\left[  X_{n-1}\right]  $)
should be backdated in the other perspective. Instead, backdating more than
half projection in one perspective, would leave less to backdate in the other,
violating the symmetry between the two perspectives.

The quantum algorithm can still be seen as a superposition of all the possible
ways of taking 50\% of $\mathcal{I}$ and all the possible results of computing
the missing information. It suffices to track each individual term of the
superposition throughout the computation. However, after each computation of
$\delta\left(  \mathbf{k},\mathbf{x}\right)  $, each history branches into
$2^{n}$ histories because of the rotation of the basis of register $X$. The
history superposition picture still explains how the quantum algorithm
exploits the advanced information, however it becomes very complex.

\section{Checking the 50\% rule on other quantum algorithms}

We check the 50\% rule on Deutsch\&Jozsa's, Simon's, and the hidden subgroup
algorithms, see Ref. $\left[  2\right]  $. Problem solving is still seen as a
game between two players. Given a set of functions $f_{\mathbf{k}}:\left\{
0,1\right\}  ^{n}\rightarrow\left\{  0,1\right\}  ^{m}$ known to both players,
the oracle chooses a function $f_{\mathbf{k}}\left(  \mathbf{x}\right)  $ and
gives to Alice a black box that, given in input a value of $\mathbf{x}$,
computes $f_{\mathbf{k}}\left(  \mathbf{x}\right)  $. Alice should find a
property of the function\ by computing $f_{\mathbf{k}}\left(  \mathbf{x}%
\right)  $\ for various values of $\mathbf{x}$.

\subsection{Deutsch\&Jozsa's algorithm}

In Deutsch\&Jozsa's algorithm, the set of functions is all the constant and
"balanced" functions (with an even number of zeroes and ones) $f_{\mathbf{k}%
}:\left\{  0,1\right\}  ^{n}\rightarrow\left\{  0,1\right\}  $. Table
(\ref{dj}) gives this set of functions for $n=2$. The string $\mathbf{k}\equiv
k_{0},k_{1},...,k_{2^{n}-1}$ is both the suffix and the table of the function
-- the sequence of function values for increasing values of the argument.
\begin{equation}%
\begin{tabular}
[c]{|c|c|c|c|c|c|c|c|c|}\hline
$\mathbf{x}$ & $\,f_{0000}\left(  \mathbf{x}\right)  $ & $f_{1111}\left(
\mathbf{x}\right)  $ & $f_{0011}\left(  \mathbf{x}\right)  $ & $f_{1100}%
\left(  \mathbf{x}\right)  $ & $f_{0101}\left(  \mathbf{x}\right)  $ &
$f_{1010}\left(  \mathbf{x}\right)  $ & $f_{0110}\left(  \mathbf{x}\right)  $
& $f_{1001}\left(  \mathbf{x}\right)  $\\\hline
00 & 0 & 1 & 0 & 1 & 0 & 1 & 0 & 1\\\hline
01 & 0 & 1 & 0 & 1 & 1 & 0 & 1 & 0\\\hline
10 & 0 & 1 & 1 & 0 & 0 & 1 & 1 & 0\\\hline
11 & 0 & 1 & 1 & 0 & 1 & 0 & 0 & 1\\\hline
\end{tabular}
\label{dj}%
\end{equation}

One should find whether the function chosen by the oracle is balanced or
constant, by computing $f_{\mathbf{k}}\left(  \mathbf{x}\right)  =f\left(
\mathbf{k},\mathbf{x}\right)  $. In the classical case this requires, in the
worst case, a number of computations of $f_{\mathbf{k}}\left(  \mathbf{x}%
\right)  $ exponential in $n$; in the quantum case one computation -- see Ref.
$\left[  7\right]  $.

For simplicity, we develop only the kernel of the quantum algorithm in Alice's
perspective. The initial state is:%

\begin{align}
&  \frac{1}{8}\left(  \left\vert 0000\right\rangle _{K}+\left\vert
1111\right\rangle _{K}+\left\vert 0011\right\rangle _{K}+\left\vert
1100\right\rangle _{K}+...\right) \label{indj}\\
&  \left(  \left\vert 00\right\rangle _{X}+\left\vert 01\right\rangle
_{X}+\left\vert 10\right\rangle _{X}+\left\vert 11\right\rangle _{X}\right)
\left(  \left\vert 0\right\rangle _{V}-\left\vert 1\right\rangle _{V}\right)
.\nonumber
\end{align}
Computing $f\left(  \mathbf{k},\mathbf{x}\right)  $ and modulo $2$\ adding the
result to the former content of $V$, yields the state of maximal entanglement:%

\begin{equation}
\frac{1}{8}\left[
\begin{array}
[c]{c}%
(\left\vert 0000\right\rangle _{K}-\left\vert 1111\right\rangle _{K}%
)(\left\vert 00\right\rangle _{X}+\left\vert 01\right\rangle _{X}+\left\vert
10\right\rangle _{X}+\left\vert 11\right\rangle _{X})+\\
(\left\vert 0011\right\rangle _{K}-\left\vert 1100\right\rangle _{K}%
)(\left\vert 00\right\rangle _{X}+\left\vert 01\right\rangle _{X}-\left\vert
10\right\rangle _{X}-\left\vert 11\right\rangle _{X})+...
\end{array}
\right]  \left(  \left\vert 0\right\rangle _{V}-\left\vert 1\right\rangle
_{V}\right)  . \label{evaluation}%
\end{equation}
Performing Hadamard on register $X$ yields:%

\begin{align}
&  \frac{1}{4}\left[  \left(  \left\vert 0000\right\rangle _{K}-\left\vert
1111\right\rangle _{K}\right)  \left\vert 00\right\rangle _{X}+(\left\vert
0011\right\rangle _{K}-\left\vert 1100\right\rangle _{K})\left\vert
10\right\rangle _{X}+....\right] \label{hdj}\\
&  \left(  \left\vert 0\right\rangle _{V}-\left\vert 1\right\rangle
_{V}\right)  .\nonumber
\end{align}
Measuring $\left[  K\right]  $ and $\left[  X\right]  $\ in the output state
(\ref{hdj}) yields the oracle's choice and the solution found by Alice: all
zeroes if the function is constant, not so if it is balanced.

We check that the quantum algorithm requires the number of oracle's queries of
a classical algorithm that knows in advance 50\% of $\mathcal{I}$. Since the
solution is a function of $\mathbf{k}$, we can define the advanced information
as any 50\% of the information about the solution contained in $\mathbf{k}$,
namely in the table of $f_{\mathbf{k}}\left(  \mathbf{x}\right)  $. If
$f_{\mathbf{k}}\left(  \mathbf{x}\right)  $ is constant, for reasons of
symmetry, the advanced information is any 50\% of the table of the function --
see table (\ref{dj}). If the function is balanced, still for reasons of
symmetry, it is any 50\% of the table that does not contain different values
of the function -- for each balanced function there are two such half tables.
In fact, the half tables that contain different values of the function already
tell that the function is balanced and thus contain 100\% of $\mathcal{I}$ .
For the \textit{good} half tables, that do not contain different values of the
function, the solution (whether the function is constant or balanced) is
always identified by computing $f_{\mathbf{k}}\left(  \mathbf{x}\right)  $ for
any value of $\mathbf{x}$\ outside the half table. Thus, both the quantum
algorithm and the advanced information classical algorithm require just one
oracle's query.

We show that the advanced information, as defined above, is backdated
projection. With respect to Grover's algorithm, we have used a different way
of representing both $\mathbf{k}$ and the advanced information. It is
instructive to retrofit Grover's algorithm in this way.

The equivalent of table (\ref{dj}) for Grover's algorithm is:%
\begin{equation}%
\begin{tabular}
[c]{|c|c|c|c|c|}\hline
$\mathbf{x}$ & $\delta\left(  00,\mathbf{x}\right)  =f_{1000}\left(
\mathbf{x}\right)  $ & $\delta\left(  01,\mathbf{x}\right)  =f_{0100}\left(
\mathbf{x}\right)  $ & $\delta\left(  10,\mathbf{x}\right)  =f_{0010}\left(
\mathbf{x}\right)  $ & $\delta\left(  11,\mathbf{x}\right)  =f_{0001}\left(
\mathbf{x}\right)  $\\\hline
00 & 1 & 0 & 0 & 0\\\hline
01 & 0 & 1 & 0 & 0\\\hline
10 & 0 & 0 & 1 & 0\\\hline
11 & 0 & 0 & 0 & 1\\\hline
\end{tabular}
\ \ . \label{gr}%
\end{equation}
The advanced information is any half table of $f_{\mathbf{k}}\left(
\mathbf{x}\right)  $ that does not contain the value 1 of the function. The
reduced density operator of register $K$, throughout the unitary part of the
algorithm in Alice's perspective, is:%
\begin{equation}
\frac{1}{2}\left(  \left\vert 1000\right\rangle _{K}+\operatorname{e}%
^{i\varphi_{01}}\left\vert 0100\right\rangle _{K}+\operatorname{e}%
^{i\varphi_{10}}\left\vert 0010\right\rangle _{K}+\operatorname{e}%
^{i\varphi_{11}}\left\vert 0001\right\rangle _{K}\right)  . \label{redo}%
\end{equation}
Let us assume that the advanced information (a good half table) is
$f_{\mathbf{k}}\left(  10\right)  =0$ and $f_{\mathbf{k}}\left(  11\right)
=0$, which means that the function chosen by the oracle is either
$f_{1000}\left(  \mathbf{x}\right)  $ or $f_{0100}\left(  \mathbf{x}\right)
$. This corresponds to projecting the reduced density operator (\ref{redo}%
)\ on $\frac{1}{\sqrt{2}}\left(  \left\vert 1000\right\rangle _{K}%
+\operatorname{e}^{i\varphi_{01}}\left\vert 0100\right\rangle _{K}\right)  $
(by the way, this projection corresponds, in the representation of section
2.3, to measuring $\left[  K_{0}\right]  $ in the output state (\ref{outputa})
and finding $k_{0}=0$). The outcome of the projection goes back unaltered to
before running the algorithm, where it becomes Alice's advanced knowledge of
the solution.

Furthermore, we can see that this advanced knowledge cannot exceed 50\% of
$\mathcal{I}$. In fact, increasing any good half table by one row, projects
the output state on the solution, leaving to the oracle nothing to backdate.

We show that the oracle's query stage of Deutsch\&Jozsa's algorithm is a
superposition of the histories of the advanced information classical
algorithm. Let us assume that the advanced information is $f\left(
\mathbf{k},00\right)  =0$ and $f\left(  \mathbf{k},01\right)  =0$, namely the
first two rows of either $\,f_{0000}\left(  \mathbf{x}\right)  $ or
$f_{0011}\left(  \mathbf{x}\right)  $ -- see table (\ref{dj}). To find the
value of $\mathbf{k}$, Alice should query the oracle with either
$\mathbf{x}=10$ or $\mathbf{x}=11$. We assume that she queries with
$\mathbf{x}=10$. If the result of the computation is $0$, \ this means that
$\mathbf{k}=0000$. This originates two histories. History \# 1: initial state
$\left\vert 0000\right\rangle _{K}\left\vert 10\right\rangle _{X}\left\vert
0\right\rangle _{V}$, state after the computation $\left\vert
0000\right\rangle _{K}\left\vert 10\right\rangle _{X}\left\vert 0\right\rangle
_{V}$. History \# 2: initial state $\left\vert 0000\right\rangle
_{K}\left\vert 10\right\rangle _{X}\left\vert 1\right\rangle _{V}$, state
after the computation $\left\vert 0000\right\rangle _{K}\left\vert
10\right\rangle _{X}\left\vert 1\right\rangle _{V}$. If the result of the
computation is $1$, \ this means that $\mathbf{k}=0011$. This originates two
histories. History \# 3: initial state $\left\vert 0011\right\rangle
_{K}\left\vert 10\right\rangle _{X}\left\vert 0\right\rangle _{V}$, state
after the computation $\left\vert 0011\right\rangle _{K}\left\vert
10\right\rangle _{X}\left\vert 1\right\rangle _{V}$. History \# 4: initial
state $\left\vert 0011\right\rangle _{K}\left\vert 10\right\rangle
_{X}\left\vert 1\right\rangle _{V}$, state after the computation $\left\vert
0011\right\rangle _{K}\left\vert 10\right\rangle _{X}\left\vert 0\right\rangle
_{V}$.\ If she queries the oracle with $\mathbf{x}=11$ instead, this
originates other 4 histories, etc.

To synthesize the quantum algorithm out of the advanced information classical
algorithm, we should choose history phases and rotation of the basis of
register $X$ in such a way that the information about the solution readable in
that register at the end of the algorithm is maximized.

\subsection{Simon's and the hidden subgroup algorithms}

The set of functions is all the $f_{\mathbf{k}}:\left\{  0,1\right\}
^{n}\rightarrow\left\{  0,1\right\}  ^{n-1}$ such that $f_{\mathbf{k}}\left(
\mathbf{x}\right)  =f_{\mathbf{k}}\left(  \mathbf{y}\right)  $ if and only if
$\mathbf{x}=\mathbf{y}$\ or $\mathbf{x}=\mathbf{y}\oplus\mathbf{h}^{\left(
\mathbf{k}\right)  }$; $\oplus$\ denotes bitwise modulo 2 addition; the bit
string $\mathbf{h}^{\left(  \mathbf{k}\right)  }\mathbf{\equiv~}h_{0}^{\left(
\mathbf{k}\right)  },h_{1}^{\left(  \mathbf{k}\right)  },...,h_{n-1}^{\left(
\mathbf{k}\right)  }$, depending on $\mathbf{k}$ and belonging to $\left\{
0,1\right\}  ^{n}$ excluded the all zeroes string, is a sort of period of the
function. Table (\ref{periodic}) gives the set of functions for $n=2$. The bit
string $\mathbf{k}$ is both the suffix and the table of the function. Since
$\mathbf{h}^{\left(  \mathbf{k}\right)  }\oplus\mathbf{h}^{\left(
\mathbf{k}\right)  }=0$, each value of the function appears exactly twice in
the table, thus 50\% of the rows plus one surely identify $\mathbf{h}^{\left(
\mathbf{k}\right)  }$.
\begin{equation}%
\begin{tabular}
[c]{|c|c|c|c|c|c|c|}\hline
& $\mathbf{h}^{\left(  0011\right)  }=01$ & $\mathbf{h}^{\left(  1100\right)
}=01$ & $\mathbf{h}^{\left(  0101\right)  }=10$ & $\mathbf{h}^{\left(
1010\right)  }=10$ & $\mathbf{h}^{\left(  0110\right)  }=11$ & $\mathbf{h}%
^{\left(  1001\right)  }=11$\\\hline
$\mathbf{x}$ & $f_{0011}\left(  \mathbf{x}\right)  $ & $f_{1100}\left(
\mathbf{x}\right)  $ & $f_{0101}\left(  \mathbf{x}\right)  $ & $f_{1010}%
\left(  \mathbf{x}\right)  $ & $f_{0110}\left(  \mathbf{x}\right)  $ &
$f_{1001}\left(  \mathbf{x}\right)  $\\\hline
00 & 0 & 1 & 0 & 1 & 0 & 1\\\hline
01 & 0 & 1 & 1 & 0 & 1 & 0\\\hline
10 & 1 & 0 & 0 & 1 & 1 & 0\\\hline
11 & 1 & 0 & 1 & 0 & 0 & 1\\\hline
\end{tabular}
\label{periodic}%
\end{equation}

The oracle chooses a function. The problem is finding the value of
$\mathbf{h}^{\left(  \mathbf{k}\right)  }$, "hidden" in the $f_{\mathbf{k}%
}\left(  \mathbf{x}\right)  $ chosen by the oracle, by computing
$f_{\mathbf{k}}\left(  \mathbf{x}\right)  $ for different values of
$\mathbf{x}$. In present knowledge, a classical algorithm requires a number of
computations of $f_{\mathbf{k}}\left(  \mathbf{x}\right)  $ exponential in
$n$. The quantum algorithm solves the hard part of this problem, namely
finding a string $\mathbf{s}_{j}^{\left(  \mathbf{k}\right)  }$
orthogonal\footnote{The modulo 2 addition of the bits of the bitwise product
of the two strings should be zero.} to $\mathbf{h}^{\left(  \mathbf{k}\right)
}$, with one computation of $f_{\mathbf{k}}\left(  \mathbf{x}\right)  $ -- see
Ref. $\left[  8\right]  $. There are $2^{n-1}$ such strings. Running the
quantum algorithm yields one of these strings at random (see further below).
The quantum algorithm is iterated until finding $n-1$ different strings. This
allows to find $\mathbf{h}^{\left(  \mathbf{k}\right)  }$ by solving a system
of modulo 2 linear equations. The black box, given $\mathbf{k}$ and
$\mathbf{x}$, computes $f_{\mathbf{k}}\left(  \mathbf{x}\right)  =f\left(
\mathbf{k},\mathbf{x}\right)  $. Register $K$\ is now $2^{n}\left(
n-1\right)  $-qubit, given that $\mathbf{k}$ is the sequence of $2^{n}$ fields
each on $n-1$\ bits.

We develop the kernel of the quantum algorithm in Alice's perspective. The
initial state, with register $V$ prepared in the all zeroes string (just one
zero for $n=2$), is:%

\begin{align}
&  \frac{1}{2\sqrt{6}}\left(  \left\vert 0011\right\rangle _{K}+\left\vert
1100\right\rangle _{K}+\left\vert 0101\right\rangle _{K}+\left\vert
1010\right\rangle _{K}+...\right) \label{insimon}\\
&  \left(  \left\vert 00\right\rangle _{X}+\left\vert 01\right\rangle
_{X}+\left\vert 10\right\rangle _{X}+\left\vert 11\right\rangle _{X}\right)
\left\vert 0\right\rangle _{V}.\nonumber
\end{align}
\ Computing $f\left(  \mathbf{k},\mathbf{x}\right)  $ changes the content of
$V$ from zero to the outcome of the computation, yielding the entangled state:%

\begin{equation}
\frac{1}{2\sqrt{6}}\left[
\begin{array}
[c]{c}%
(\left\vert 0011\right\rangle _{K}+\left\vert 1100\right\rangle _{K})\left[
(\left\vert 00\right\rangle _{X}+\left\vert 01\right\rangle _{X})\left\vert
0\right\rangle _{V}+(\left\vert 10\right\rangle _{X}+\left\vert
11\right\rangle _{X})\left\vert 1\right\rangle _{V}\right]  +\\
(\left\vert 0101\right\rangle _{K}+\left\vert 1010\right\rangle _{K})\left[
(\left\vert 00\right\rangle _{X}+\left\vert 10\right\rangle _{X})\left\vert
0\right\rangle _{V}+(\left\vert 01\right\rangle _{X}+\left\vert
11\right\rangle _{X})\left\vert 1\right\rangle _{V}\right]  +...
\end{array}
\right]  . \label{second}%
\end{equation}
Performing Hadamard on $X$ yields:%

\begin{equation}
\frac{1}{2\sqrt{6}}\left[
\begin{array}
[c]{c}%
(\left\vert 0011\right\rangle _{K}+\left\vert 1100\right\rangle _{K})\left[
(\left\vert 00\right\rangle _{X}+\left\vert 10\right\rangle _{X})\left\vert
0\right\rangle _{V}+(\left\vert 00\right\rangle _{X}-\left\vert
10\right\rangle _{X})\left\vert 1\right\rangle _{V}\right]  +\\
(\left\vert 0101\right\rangle _{K}+\left\vert 1010\right\rangle _{K})\left[
(\left\vert 00\right\rangle _{X}+\left\vert 01\right\rangle _{X})\left\vert
0\right\rangle _{V}+(\left\vert 00\right\rangle _{X}-\left\vert
01\right\rangle _{X})\left\vert 1\right\rangle _{V}\right]  +...
\end{array}
\right]  , \label{hsimon}%
\end{equation}
where, for each value of $\mathbf{k}$, register $X$ (no matter the content of
$V$) hosts even weighted\ superpositions of the $2^{n-1}$ strings
$\mathbf{s}_{j}^{\left(  \mathbf{k}\right)  }$ orthogonal to $\mathbf{h}%
^{\left(  \mathbf{k}\right)  }$. By measuring $\left[  K\right]  $\ and
$\left[  X\right]  $ in state (\ref{hsimon}), we obtain at random the oracle's
choice $\mathbf{k}$ and one of the $\mathbf{s}_{j}^{\left(  \mathbf{k}\right)
}$.

We leave $K$ in its after-measurement state, thus fixing $\mathbf{k}$, and
iterate the "right part" of the algorithm (preparation of registers $X$\ and
$V$, computation of $f\left(  \mathbf{k},\mathbf{x}\right)  $, and measurement
of $\left[  X\right]  $) until obtaining $n-1$ different $\mathbf{s}%
_{j}^{\left(  \mathbf{k}\right)  }$.

We check that the quantum algorithm requires the number of oracle's queries of
a classical algorithm that knows in advance 50\% of $\mathcal{I}$. Any
$\mathbf{s}_{j}^{\left(  \mathbf{k}\right)  }$ is a solution of the problem
addressed by the quantum part of Simon's algorithm. The advanced information
is any 50\% of the information about the solution contained in $\mathbf{k}$.
For reasons of symmetry, this is any 50\% of the table of the function that
does not contain the same value of the function twice. In fact, the half
tables that contain a same value twice already specify the value of
$\mathbf{h}^{\left(  \mathbf{k}\right)  }$ and thus the value of any
$\mathbf{s}_{j}^{\left(  \mathbf{k}\right)  }$. For the half tables that do
not contain the same value of the function twice, the solution is always
identified by computing $f\left(  \mathbf{k},\mathbf{x}\right)  $ for any
value of $\mathbf{x}$\ outside the half table. The new value of the function
is necessarily a value already present in the half table, which identifies
$\mathbf{h}^{\left(  \mathbf{k}\right)  }$ and all the $\mathbf{s}%
_{j}^{\left(  \mathbf{k}\right)  }$. Thus, both the quantum algorithm and the
advanced information classical algorithm require just one oracle's query.

As in section 3.1, the above defined advanced information is backdated
projection on the solution.

We show that the oracle's query stage of the quantum algorithm is a
superposition of the histories of the advanced information classical
algorithm.\ For example, let us assume that the advanced information is
$f\left(  \mathbf{k},00\right)  =0$ and $f\left(  \mathbf{k},11\right)  =1$,
namely the first and last row of either $f_{0011}\left(  \mathbf{x}\right)  $
or $f_{0101}\left(  \mathbf{x}\right)  $ -- see table (\ref{periodic}). To
find the value of $\mathbf{k}$, Alice should query the oracle with either
$\mathbf{x}=01$ or $\mathbf{x}=10$. We assume that she queries with
$\mathbf{x}=01$. If the result of the computation is $0$, this means that
$\mathbf{k}=0011$. This originates two histories. History \# 1: initial state
$\left\vert 0011\right\rangle _{K}\left\vert 01\right\rangle _{X}\left\vert
0\right\rangle _{V}$, state after the computation $\left\vert
0011\right\rangle _{K}\left\vert 01\right\rangle _{X}\left\vert 0\right\rangle
_{V}$. History \# 2: initial state $\left\vert 0011\right\rangle
_{K}\left\vert 01\right\rangle _{X}\left\vert 1\right\rangle _{V}$, state
after the computation $\left\vert 0011\right\rangle _{K}\left\vert
01\right\rangle _{X}\left\vert 1\right\rangle _{V}$. If the result of the
computation is $1$, this means that $\mathbf{k}=0101$. This originates two
histories. History \# 3: initial state $\left\vert 0101\right\rangle
_{K}\left\vert 01\right\rangle _{X}\left\vert 0\right\rangle _{V}$, state
after the computation $\left\vert 0101\right\rangle _{K}\left\vert
01\right\rangle _{X}\left\vert 1\right\rangle _{V}$. History \# 3: initial
state $\left\vert 0101\right\rangle _{K}\left\vert 01\right\rangle
_{X}\left\vert 1\right\rangle _{V}$, state after the computation $\left\vert
0101\right\rangle _{K}\left\vert 01\right\rangle _{X}\left\vert 0\right\rangle
_{V}$.\ If she queries the oracle with $\mathbf{x}=10$ instead, this
originates other 4 histories, etc.

To synthesize the quantum algorithm out of the advanced information classical
algorithm, we should choose history phases and rotation of the basis of
register $X$ in such a way that the information about the solution readable in
that register at the end of the algorithm is maximized.

The 50\% rule also applies to the generalized Simon's problem and to the
hidden subgroup problem. In fact the corresponding algorithms are essentially
the same as the algorithm that solves Simon's problem. In the hidden subgroup
problem, the set of functions $f_{\mathbf{k}}:G\rightarrow W$ map a group $G$
to some finite set $W$\ with the property that there exists some subgroup
$S\leq G$ such that for any $x,y\in G$, $f_{\mathbf{k}}\left(  x\right)
=f_{\mathbf{k}}\left(  y\right)  $ if and only if $x+S=y+S$. The problem is to
find the hidden subgroup $S$ by computing $f_{\mathbf{k}}\left(  x\right)  $
for various values of $x$. Now, a large variety of quantum problems can be
re-formulated in terms of the hidden subgroup problem $\left[  9\right]  $.
Among these we find: Deutsch's problem, Bernstein\&Vazirani problem, finding
orders, finding the period of a function (thus the problem solved by the
quantum part of Shor's factorization algorithm), discrete logarithms in any
group, hidden linear functions, self shift equivalent polynomials, Abelian
stabilizer problem, graph automorphism problem.

\section{Applying the 50\% rule to the search of quantum speed ups}

In hindsight, the quantum algorithms examined are skillfully designed around
the 50\% rule. In unstructured data base search, the advanced knowledge of
50\% of $\mathcal{I}$ yields a quadratic speed up, given that the number of
oracle's queries goes from $\operatorname{O}\left(  2^{n}\right)  $ to
$\operatorname{O}\left(  2^{n/2}\right)  $. Thus, the possibility of a
quadratic speed up is established by the 50\% rule, one does not need to know
Grover's algorithm. Similarly, in the structured algorithms that yield an
exponential speed up, the problem is chosen in such a way that, if one knows
in advance 50\% of $\mathcal{I}$, computing $f_{\mathbf{k}}\left(
\mathbf{x}\right)  $ for a single value of $\mathbf{x}$\ outside the advanced
information yields the solution. Thus, the possibility of an exponential speed
up is established by the 50\% rule before knowing the quantum algorithm.

One way of searching for new quantum speed ups is thus looking for problems
solvable with a single computation of $f_{\mathbf{k}}\left(  \mathbf{x}%
\right)  $\ once that 50\% of $\mathcal{I}$ is known. We provide an example --
see also Ref. $\left[  2\right]  $. The set of functions is the $4!$%
\ functions $f_{\mathbf{k}}:\left\{  0,1\right\}  ^{2}\rightarrow\left\{
0,1\right\}  ^{2}$ such that the sequence of function values is a permutation
of the values of the argument -- see table (\ref{perm}).%

\begin{equation}%
\begin{tabular}
[c]{|c|c|c|c|c|}\hline
$\mathbf{x}$ & $f_{00011110}\left(  \mathbf{x}\right)  $ & $f_{00110110}%
\left(  \mathbf{x}\right)  $ & $f_{00011011}\left(  \mathbf{x}\right)  $ &
$...$\\\hline
00 & 00 & 00 & 00 & ...\\\hline
01 & 01 & 11 & 01 & ...\\\hline
10 & 11 & 01 & 10 & ...\\\hline
11 & 10 & 10 & 11 & ...\\\hline
\end{tabular}
\label{perm}%
\end{equation}
We have chosen this set because, if we know 50\% of the rows of one table, we
can identify the corresponding $\mathbf{k}$\ with a single computation of
$f_{\mathbf{k}}\left(  \mathbf{x}\right)  $, for any value of $\mathbf{x}$
outside the advanced information. Without advanced information, three
computations of $f_{\mathbf{k}}\left(  \mathbf{x}\right)  $ are required. Thus
there is room for a speed up in terms of number of oracle's queries. We build
a quantum algorithm over this possibility. Register $K$\ is $8$ qubits,
registers $X$ is $2$ qubits, and register $V$ is $2$\ qubits, denoted $V_{0}$
and $V_{1}$. The first (second) bit of the result of the computation of
$f_{\mathbf{k}}\left(  \mathbf{x}\right)  =f\left(  \mathbf{k},\mathbf{x}%
\right)  $ is modulo $2$ added to the former content of $V_{0}$ ($V_{1}$). The
initial state is%

\begin{align*}
&  \frac{1}{8\sqrt{6}}\left(  \left\vert 00011110\right\rangle _{K}+\left\vert
00110110\right\rangle _{K}+\left\vert 00011011\right\rangle _{K}...\right) \\
&  \left(  \left\vert 00\right\rangle _{X}+\left\vert 01\right\rangle
_{X}+\left\vert 10\right\rangle _{X}+\left\vert 11\right\rangle _{X}\right)
\left(  \left\vert 0\right\rangle _{V_{0}}-\left\vert 1\right\rangle _{V_{0}%
}\right)  \left(  \left\vert 0\right\rangle _{V_{1}}-\left\vert 1\right\rangle
_{V_{1}}\right)  .
\end{align*}
Performing one computation of $f\left(  \mathbf{k},\mathbf{x}\right)  $ then
Hadamard on $X$, yields%

\begin{align*}
&  \frac{1}{4\sqrt{6}}\left[  \left(  \left\vert 00011110\right\rangle
_{K}+~...\right)  \left\vert 01\right\rangle _{X}+\left(  \left\vert
00110110\right\rangle _{K}+~...\right)  \left\vert 10\right\rangle
_{X}+\left(  \left\vert 00011011\right\rangle _{K}+~...\right)  \left\vert
11\right\rangle _{X}\right] \\
&  \left(  \left\vert 0\right\rangle _{V_{0}}-\left\vert 1\right\rangle
_{V_{0}}\right)  \left(  \left\vert 0\right\rangle _{V_{1}}-\left\vert
1\right\rangle _{V_{1}}\right)  ,
\end{align*}
an entangled state where three orthogonal states of $K$ (each a superposition
of $8$ values of $\mathbf{k}$, corresponding to a partition of the set of $24$
functions) are correlated with, respectively, $\left\vert 01\right\rangle
_{X},~\left\vert 10\right\rangle _{X},~$and $\left\vert 11\right\rangle _{X}$.
Measuring $\left[  X\right]  $ in the above state tells which of the three
partitions the function belongs to. In the case of a classical algorithm,
identifying the partition requires three computations of $f\left(
\mathbf{k},\mathbf{x}\right)  $, as readily checked. There is thus a quantum
speed up.

With the 50\% rule, one can figure out any number of these speed ups in terms
of number of oracle's queries. Thus, this rule provides a playground for
studying the engineering of quantum algorithms.

\section{Conclusions}

Let $\mathcal{I}$\ be the information acquired by reading the solution of the
problem. The 50\% rule establishes that a quantum algorithm\ requires the
number of oracle's queries of a classical algorithm that knows in advance 50\%
of $\mathcal{I}$\ . The advanced knowledge of the solution is due to
backdating, to before running the algorithm, a time symmetric part of the
final projection on the solution. The computation performed by the quantum
algorithm is a superposition of classical computations that exploit the
advanced knowledge of the solution to reach the solution with fewer oracle's
queries. We have checked that the rule holds for a variety of quantum
algorithms yielding both quadratic and exponential speed up.

This article should be considered work in progress. For example, it would be
desirable to check the 50\% rule on other known quantum algorithms and to
demonstrate that the rule holds in a more general way, for example for the
generic quantum computation network.

This rule would have an important practical consequence: the speed up in terms
of number of oracle's queries comes from comparing two classical algorithms,
with and without advanced information. This allows to characterize the
problems solvable with a quantum speed up in an entirely computer science
framework, with no physics involved. By way of exemplification, we have
produced a new quantum speed up on the basis of the 50\% rule.

The possibility that quantum algorithms use backdated information about the
solution they will find in the future to reduce the number of operations
required to find the solution, involves a causality loop that could be
interesting also outside quantum computation.

\subsection*{Acknowledgements}

The author thanks Pablo Arrighi, Vint Cerf, Artur Ekert, David Finkelstein,
Hartmut Neven, Daniel Sheehan, and Henry Stapp for useful discussions, Scott
Aaronson and Charles Stromeyer for useful comments.

\subsection*{Bibliography}

$\left[  1\right]  $\ D. Gross, S. T. Flammia, and J. Eisert, Phys. Rev. Lett.
\textbf{102} (19) (2009).

$\left[  2\right]  $ G. Castagnoli, Int. J. Theor. Phys.,vol. 48 issue 12,
3383 (2009).

$\left[  3\right]  $\ G. Castagnoli, Int. J. Theor. Phys. vol. 48 issue 8,
2412 (2009).

$\left[  4\right]  $\ G. Castagnoli and D. Finkelstein, Proc. Roy. Soc. Lond.
A \textbf{457}, 1799 . arXiv:quant-ph/0010081 v1 (2001).

$\left[  5\right]  $\ L. K. Grover, Proc. 28th Ann. ACM Symp. Theory of
Computing (1996).

$\left[  6\right]  $\ \ C. Rovelli, Int. J. Theor. Phys. \textbf{35}, 1637 (1996).

$\left[  7\right]  $\ D. Deutsch and R. Jozsa, Proc. Roy. Soc. (Lond.) A,
\textbf{439}, 553 (1992).

$\left[  8\right]  $ D. Simon, Proc. 35th Ann. Symp. on Foundations of Comp.
Sci.,\textit{\ }116 (1994).

$\left[  9\right]  $ P. Kaye, R. Laflamme, and M. Mosca, \textit{An
introduction to Quantum Computing}, Oxford University Press, 146 (2007).

\end{document}